\documentclass[aps,prl,twocolumn,showpacs]{revtex4}
\usepackage{graphicx}

\begin{document}
\author{Joel Koplik}
\email{koplik@sci.ccny.cuny.edu}
\affiliation{Benjamin Levich Institute and Department of Physics,
City College of the City University of New York, New York, NY 10031}
\author{Jayanth R.\ Banavar}
\email{jayanth@phys.psu.edu}
\affiliation{Department of Physics, The Pennsylvania State University,
University Park, PA 16802}

\date{\today}
\title{Slip, immiscibility and boundary conditions at the liquid-liquid 
interface}
\begin{abstract}
The conventional boundary conditions at the interface between two flowing
liquids include continuity of the tangential velocity.  We have tested this
assumption with molecular dynamics simulations of Couette and Poiseuille
flows of two-layered liquid systems, with various molecular structures and
interactions.  When the total liquid density near the interface drops
significantly compared to the bulk values, the tangential velocity varies
very rapidly there, and would appear discontinuous at continuum resolution.
The value of this apparent slip is given by a Navier boundary condition.
\end{abstract}
\maketitle

Recent studies of the nanoscale behavior of flowing fluids have reinvigorated
interest in the nature and validity of the boundary conditions which
accompany the Navier-Stokes equations.  The velocity condition at a
solid-liquid interface, and the possibility of slip there, has been a
particular focus \cite{lauga} due to its relevance in possible ``lab on a chip''
and other devices\cite{gc,ssa}.  
At a liquid-liquid interface the conventional boundary condition is also
no-slip.  An obvious physical argument is that the interface between the two
liquids is actually a region whose thickness is at least a few molecular
diameters, where molecules of both materials are present and interacting with
each other.  It is difficult to imagine how two intermixed dense liquids 
could maintain distinct molecular speeds, and one expects a single velocity 
for both liquids
in the interface, and that this velocity would vary smoothly in moving from
the interface into either bulk region as the species concentrations change
gradually.  In the light of the examples of solid-liquid slip cited above,
this argument might fail when interfacial mixing is poor and the molecules
of different species are spatially separated.  For simple liquids, we are
not aware of any experimental measurements or systematic computational
studies of liquid-liquid slip at all, although for polymer melts there is
by now convincing indirect \cite{zm} and direct \cite{lam} evidence for slip.
The former study is based on the interpretation of measurements of pressure 
drop {\em vs}. shear rate in extrusion, and the latter on confocal microscopic 
observation with a spatial resolution of about 10$\mu$m.  At the
molecular scale, there are MD simulations for model polymers \cite{br}
and self-consistent field theory calculations \cite{lo} which find slip,
but no direct experimental results.

To investigate the question of liquid-liquid slip on a fundamental
microscopic basis, we have conducted molecular dynamics (MD) simulations of 
the Couette and Poiseuille flows of two-layered immiscible liquid systems 
for a number of simple choices of interactions and molecular architecture.
Standard MD techniques \cite{at,kb95} are used, and the computational
details are similar to those of Refs.~\cite{ckb}.  The basic interatomic
potential of Lennard-Jones form, $V_{ij}(r)=4\epsilon
\left[(r/\sigma)^{-12} -A_{ij} (r/\sigma)^{-6}\right]$, where $r$ is the
interatomic separation, $\sigma$ is roughly the size of the repulsive
core, of order a few Angstroms, $\epsilon$ is the strength of the potential 
and $A_{ij}=A_{ji}$ is a dimensionless parameter that controls the
attraction between atoms of atomic species $i$ and $j$.  Numerical results 
are expressed in terms of the length scale $\sigma$ (a few Angstroms), the 
atomic mass $m$, and a time scale $\tau=\sigma(m/\epsilon)^{1/2}$, 
a few picoseconds.  Temperature is controlled by a
Nos\'e-Hoover thermostat, and the atoms are in many cases grouped into 
flexible chain molecules using a FENE potential $V_{\rm FENE}(r)=-(k/2)
\ln\left[1-{r^2/r_0^2}\right]$ with maximum bond length
$r_0=1.5\sigma$ and spring constant $k=30\epsilon/\sigma^2$.  The liquids
are confined between solid walls, each made of a layer of fcc unit cells
whose atoms are tethered to lattice sites with a stiff linear spring.
Periodic boundary conditions are applied in the two lateral directions.
Couette flow is achieved by translating the upper and lower 
wall tether sites at constant velocity $\pm U$, and Poiseuille flow results 
from applying an acceleration $g$ parallel to the walls to each liquid atom. 

\begin{figure}
\includegraphics[width=0.8\linewidth]{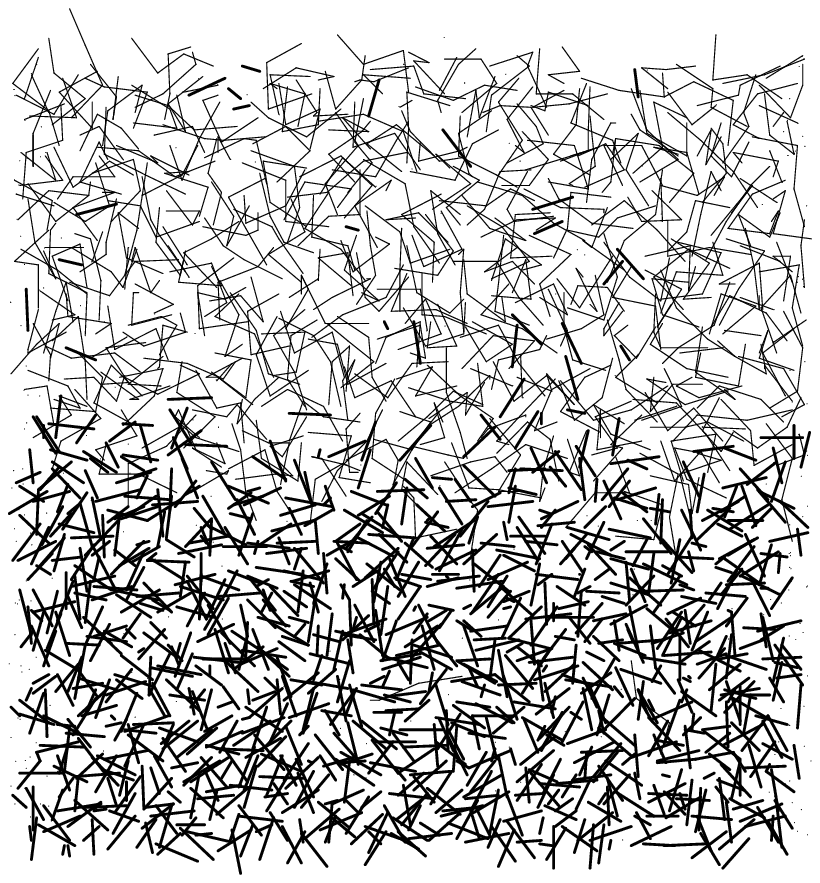}
\end{figure}
\begin{figure}
\includegraphics[width=0.8\linewidth]{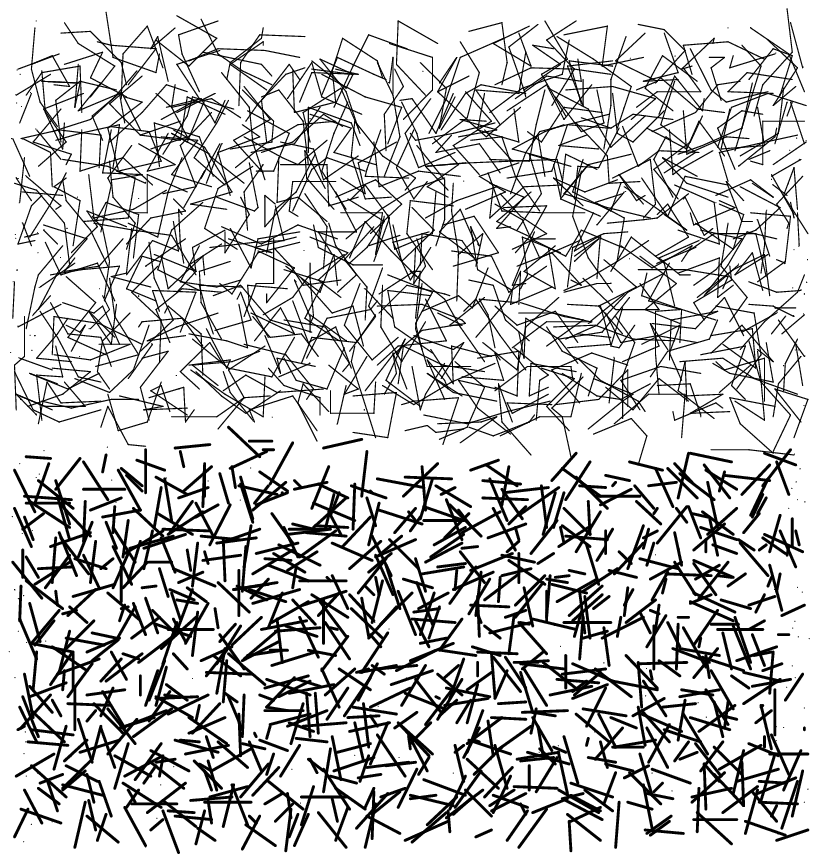}
\caption{\label{int} Interfacial region for the (2,4) system for miscible
(top) and immiscible (bottom) interactions. The figure shows a three 
dimensional slab centered on the interface, as viewed from a long distance. 
Molecules are represented by the line 
segments joining the atomic centers, using thick/thin lines for dimer/tetramer 
molecules. The width of the system is 17.1$\sigma$, and the instantaneous
gap between the upper and lower liquids in the immisicble case is
0.5-1.0$\sigma$. }
\end{figure}

The different simulated systems are characterized by the interaction
coefficients $A_{ij}$ and the lengths $\ell_1$ and $\ell_2$ -- the number 
of atoms per chain -- of the two species of liquid molecule.  The
interactions are either immiscible, with $A_{12}=0$ and all other
$A_{ij}=1$, or partially miscible as given by the Lorentz-Berthelot 
combination rules \cite{at}, with $A_{11}=5/4$, $A_{22}=3/4$, and $A_{12}=
\sqrt{A_{11}\,A_{22}}=0.97$.  Although the interactions are somewhat simplified 
as compared with realistic molecules, these systems exhibit a sufficient 
variety of behaviors to identify some trends.  We have examined systems 
with both types of interactions, for the cases $(l_1,l_2)=(1,1)$, (2,4) and 
(4,16).  In the first two cases, the simulated system consists 
of 4000 atoms of each fluid, and 576 solid atoms in each wall, and has
length $17.1\sigma$ in the flow and neutral directions and $34.2\sigma$ 
between walls;  the third system was twice as large in each dimension and has
eight times the number of atoms.  The simulated Reynolds numbers are 
$O(10^{-2}-1)$, and the Deborah number based on the characteristic atomic 
time $\tau$ is ${\rm De}= \dot{\gamma}\tau = O(10^{-2})$. 
We will discuss the results for the prototypical (2,4) case in some detail. 
Numerical results are summarized in Table I below.

\begin{figure}
\includegraphics[width=0.4\linewidth]{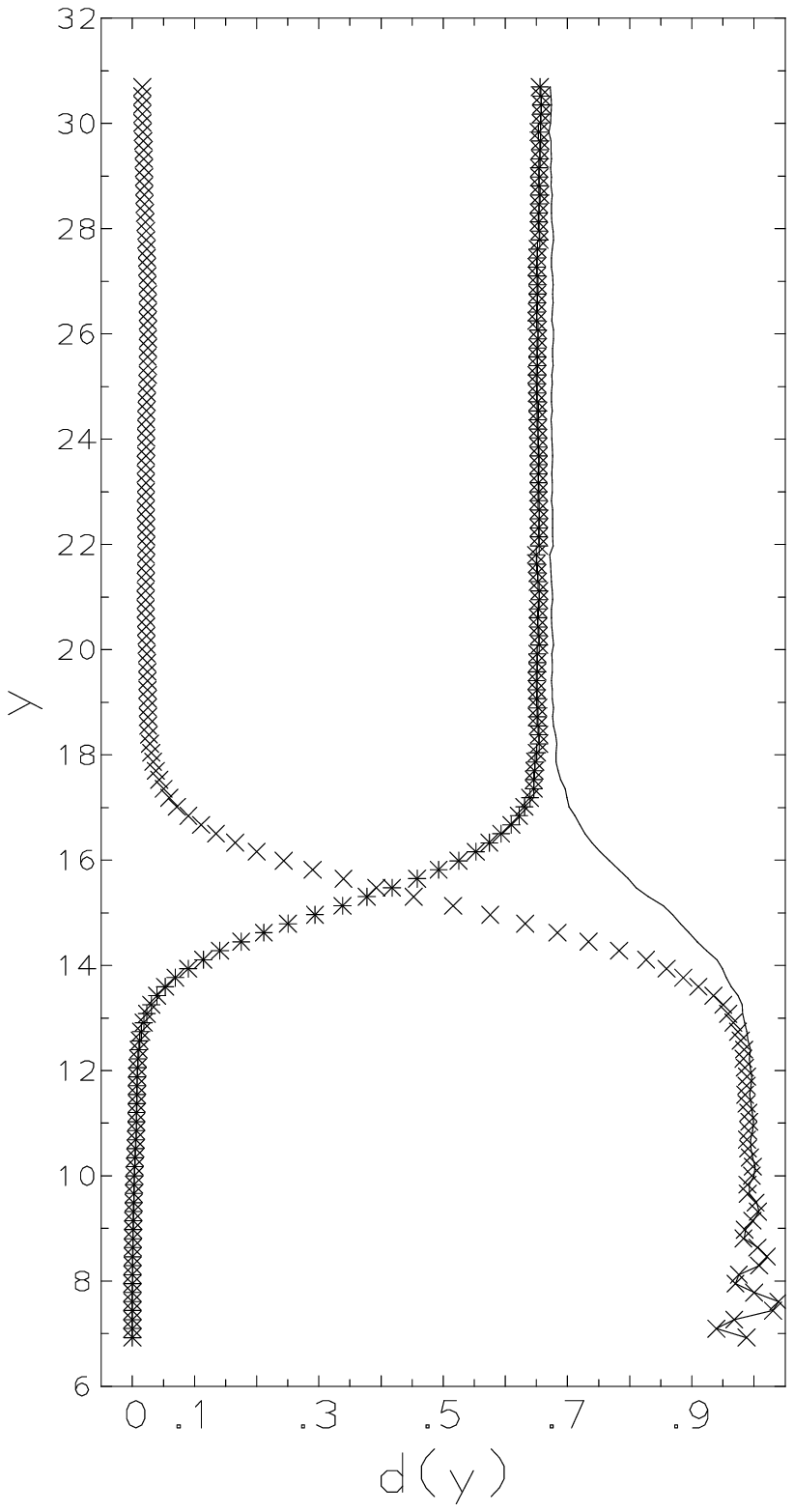}
\includegraphics[width=0.4\linewidth]{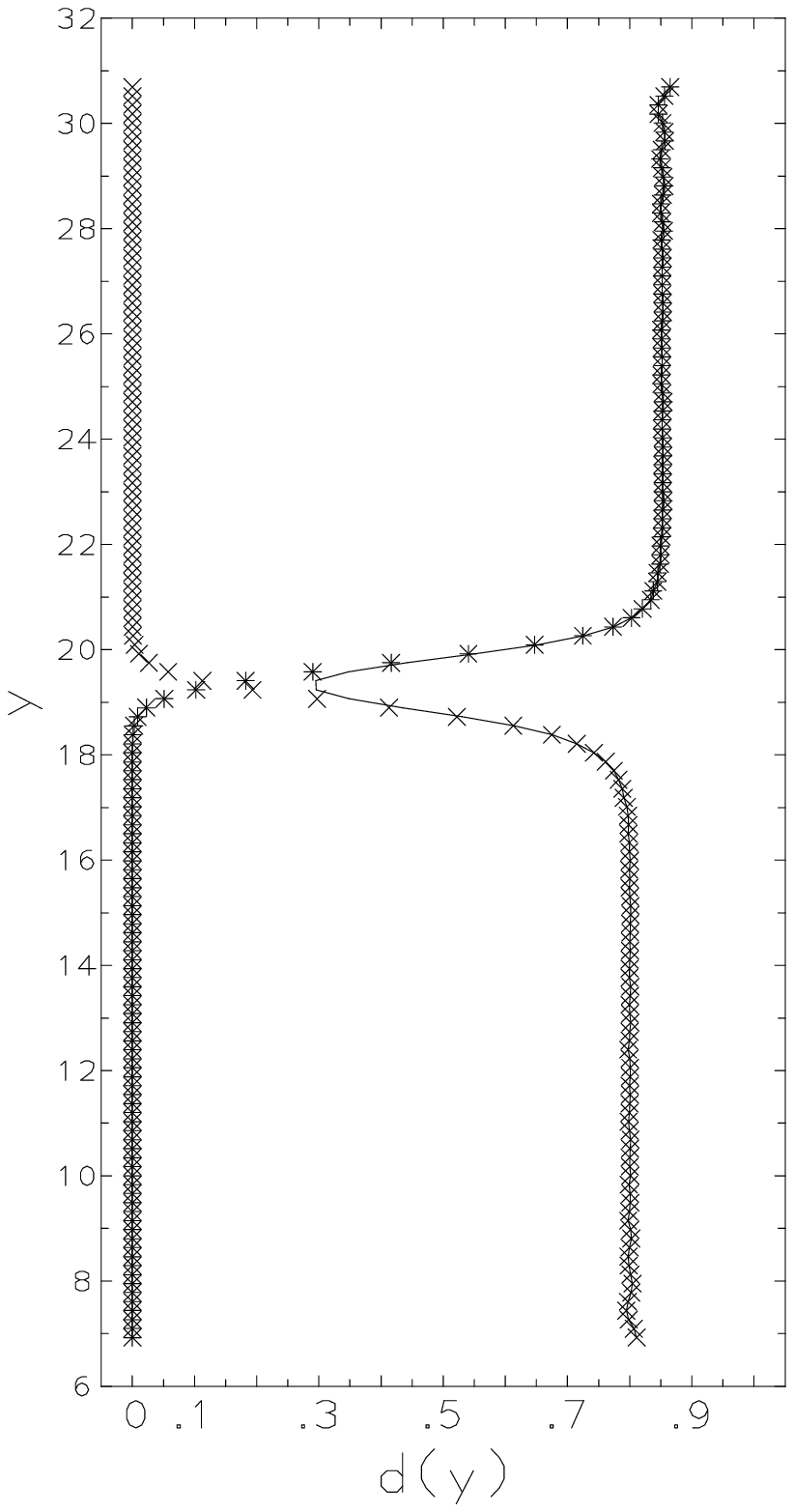}
\vspace*{0.2in}
\includegraphics[width=0.8\linewidth]{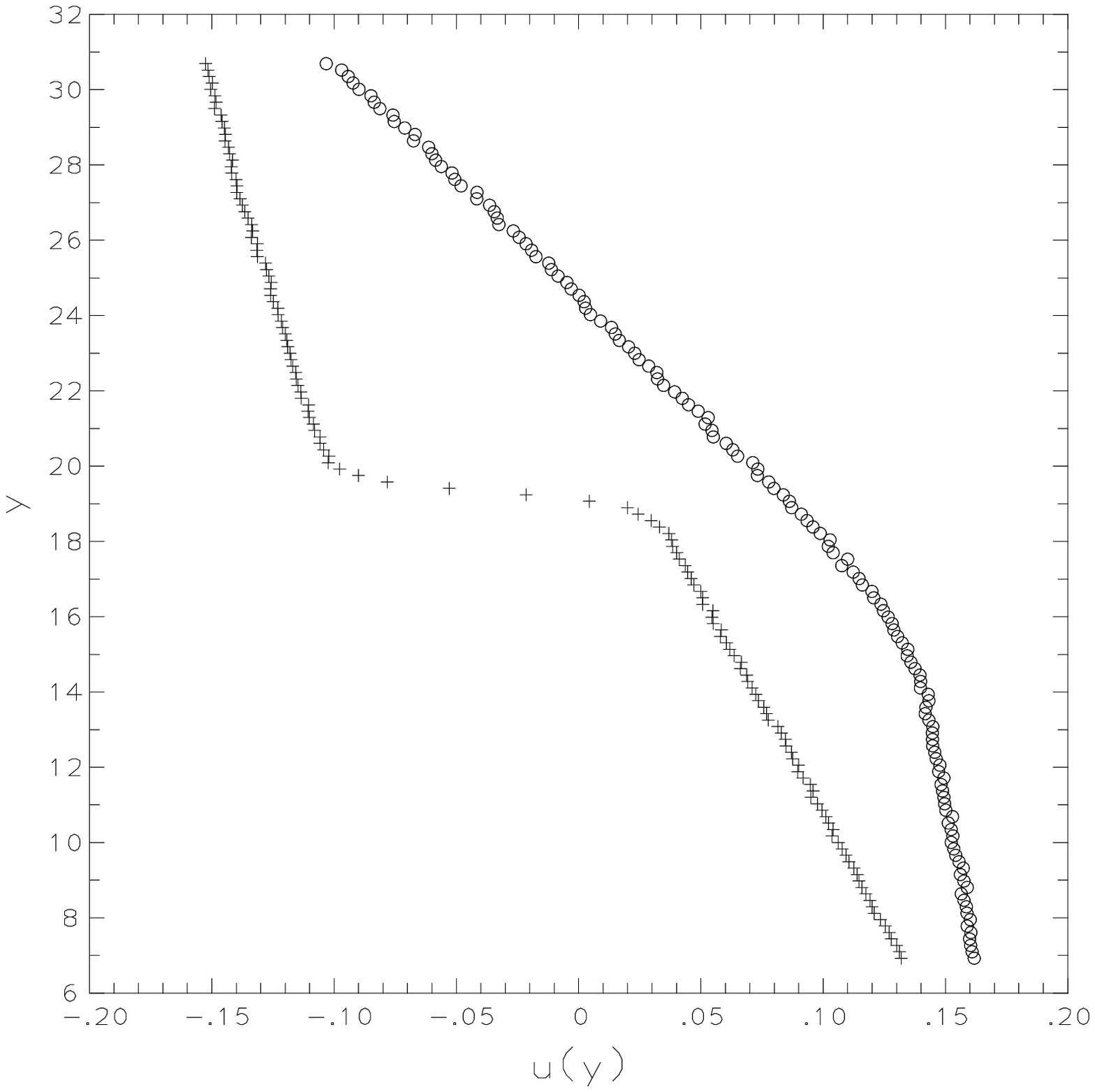}
\caption{\label{dens} Density and velocity profiles in Couette flow for
(2,4) systems. The coordinate $y$ runs normal to the interface, and the
profiles average over the other two directions.  
In the density profiles, miscible (left) and immiscible (right),
the ($\times$) symbols refer to the dimers and ($*$) to the tetramers, 
while the continuous curve is the total liquid density.  In the velocity and 
other plots below, points labeled (o) and (+) refer to the miscible and 
immiscible systems, respectively.  }
\end{figure}

\begin{figure}
\includegraphics[width=0.8\linewidth]{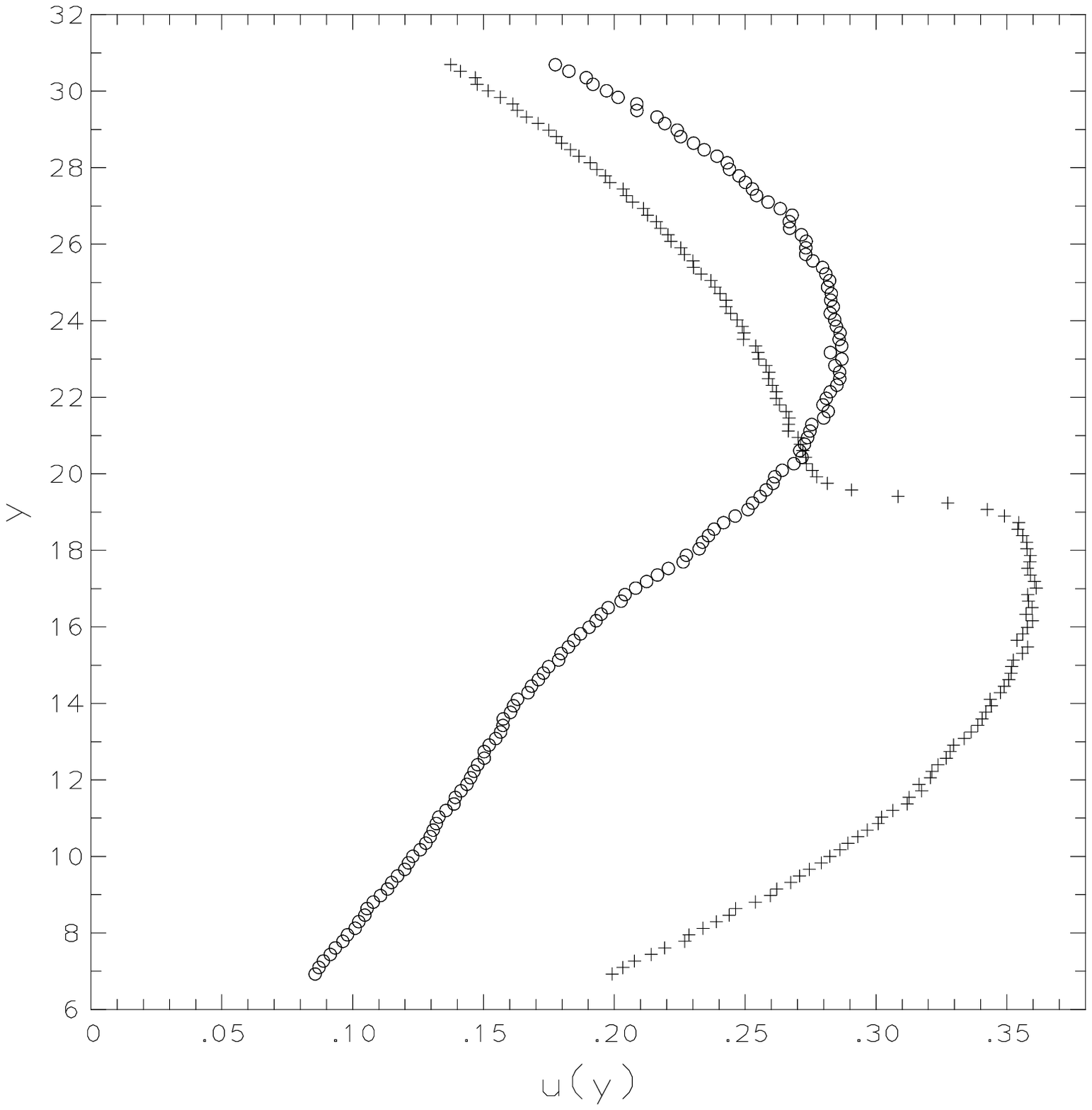}
\includegraphics[width=0.8\linewidth]{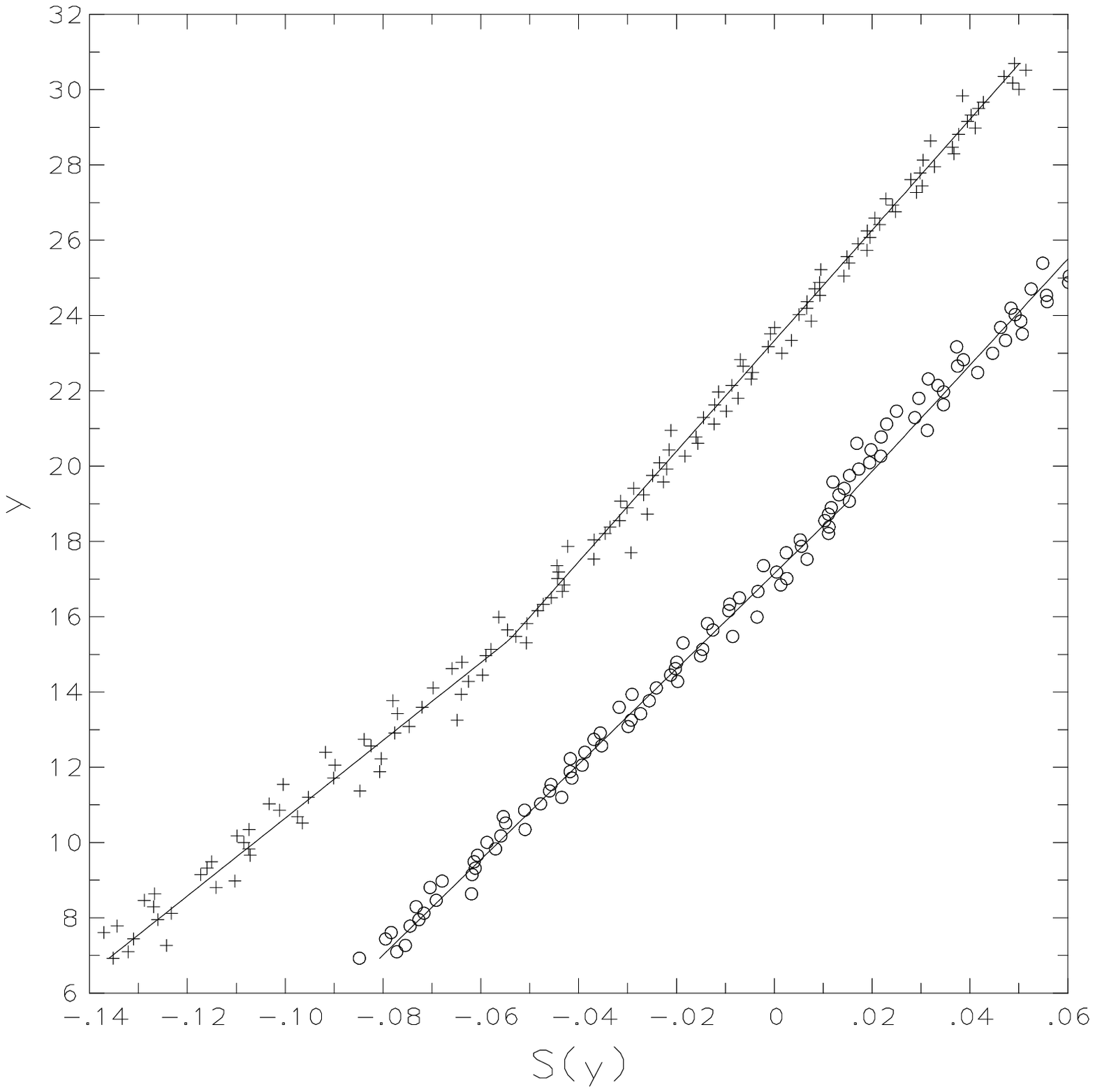}
\caption{\label{pois} Velocity and shear stress profiles in Poiseuille flow
for the (2,4) systems, with points marked by (o) and (+) miscible and 
immiscible respectively. In the stress plot, the straight lines are
a linear fit to each liquid region. } 
\end{figure}

A crucial feature of these two-liquid systems is the microscopic
structure of the interface, and in Fig.~\ref{int} we show a snapshot of
the atoms in this region for the miscible and immiscible cases.  In
the miscible case atoms of the two molecules attract each other, so an overlap
region separates the bulk liquids, whereas in the immiscible case the two
types of atom repel each other, and there is an open gap.  The time-averaged 
density profiles in Fig.~\ref{dens} reflect this behavior:  in the miscible
case the density varies monotonically from one bulk value to the other 
whereas the immiscible case shows a substantial dip in density in the 
interfacial region.  A quantitative measure of this density dip used below 
is the the difference between the mean of the two bulk densities and the 
density at the interface, relative to the mean, $\delta = 0.66$ in this
case.  In Couette flow, we see that the velocity
profile for the miscible system consists of two straight segments with
different slopes (reflecting the different viscosities of the two liquids) 
with a rounded transition located at the position of the interface.
The shear stress (not shown) has a constant value throughout both liquids.
In Poiseuille flow for the miscible system, the density profile is
essentially unchanged, while the velocity profile
Fig.~\ref{pois} corresponds to two distinct parabolas with a smooth 
transition, and the shear stress has two straight segments of different 
slope (reflecting the different liquid densities) which join smoothly 
to produce a continuous function of position.

In the immiscible case, while the velocity profiles are again
continuous functions, they exhibit a very rapid transition in traversing the 
interface in both flows, while the shear stress has the same qualitative 
features as in the previous case.  Note that 
in obtaining these density, velocity and stress profiles, we divide the
region between the walls into {\em very} narrow slabs of thickness $0.17\sigma$
parallel to the interface 
and average over a 5000$\tau$ time interval.  Most conceivable experiments
and all continuum modeling will not have the sub-Angstrom spatial resolution 
of these simulations, and the smoothed step in the velocity field in the
immiscible case would appear to be a discontinuity, which we would describe 
precisely as ``apparent velocity slip.''

\begin{table}
\begin{center}
\begin{tabular}{| l | l || c | c | c | c|} \hline
System & Flow & $\Delta u$ & $S$ & $\alpha$ & $\delta$ \\ \hline\hline
(1,1)  & 0.001 C& 0.0029& 0.0011& 2.6 & 0.49 \\ \cline{2-6}
       & 0.05 C & 0.014 & 0.0059& 2.4 & 0.49 \\ \cline{2-6}
       & 0.1 C  & 0.031 & 0.012 & 2.6 & 0.49 \\ \cline{2-6}
       & 0.2 C  & 0.050 & 0.020 & 2.5 & 0.49 \\ \cline{2-6}
       & 0.01 P & 0.0   & 0.0   & --  & 0.49 \\ \hline
(2,4)  & 0.05 C & 0.038 & 0.0063& 6.0 & 0.66 \\ \cline{2-6}
       & 0.1 C  & 0.070 & 0.012 & 5.8 & 0.66 \\ \cline{2-6}
       & 0.2 C  & 0.13  & 0.021 & 6.2 & 0.66 \\ \cline{2-6}
       & 0.01 P & 0.071 & 0.012 & 5.9 & 0.66 \\ \cline{2-6}
       & 0.02 P & 0.15  & 0.025 & 6.0 & 0.66 \\ \hline
(4,16) & 0.1 C  & 0.032 & 0.0099& 3.2 & 0.46 \\ \cline{2-6}
       & 0.2 C  & 0.070 & 0.021 & 3.3 & 0.46 \\ \cline{2-6}
       & 0.01 P & 0.15  & 0.049 & 3.0 & 0.46 \\ \cline{2-6}
       & 0.02 P & 0.22  & 0.70  & 3.1 & 0.46 \\ \hline
\end{tabular}
\end{center}
\caption{\label{tableI} Numerical results for slip.  The notation is that
$(l_1,l_2)$ refers to
a liquid made of flexible chains of length $l_1$ in contact with a second
liquid of chains of length $l_2$, with interactions either of the immiscible
or miscible (LB) type.  ``0.1 C'' means Couette flow with wall velocities
$\pm$0.1, and "0.01 P" means gravity driven Poiseuille flow with
acceleration 0.01. $\Delta u$, $S$, $\alpha$ and $\delta$ are the apaprent
slip, shear stress at the interface, Navier coefficient and relative density
dip, respectively.  All entries are in MD units. }
\end{table}

It remains to characterize the velocity discontinuity in terms of a boundary
condition suitable for continuum calculations.  Following the history of
the no-slip condition \cite{lauga}, simple plausibility, and the results of 
Zhao and Macosko \cite{zm} for polymer systems, we consider the Navier
condition $\Delta u = \alpha\, S$, where $S$ is the shear stress at 
the position of the interface, and $\alpha$ is a slip coefficient that
depends on the nature of the two liquids present.  ($\alpha$ is the inverse
of the coefficient $\beta$ introduced in \cite{zm}.)  
In the (1,1) immiscible system, the two liquids are identical except for
their mutual repulsion and have equal viscosities and densities, so that the
interface lies exactly in the middle of the channel.  In Poiseuille flow, 
the shear stress then vanishes at the interface, and the Navier condition
predicts no velocity discontinuity, exactly as seen in the simulations.

In Table I, we evaluate the slip coefficient from the MD data in the 
various cases simulated.  The key feature of the Table is
the approximately constant value of $\alpha$ obtained for each liquid pair,
independent of the flow configuration and the value of the driving force.
The numerical values have not been determined with very high precision,
partly due to statistical fluctuations in the shear stress, and partly due to
uncertainties in extrapolating across the interfacial region, but the trend
is clear.  At sufficiently high shear rates, non-Newtonian effects would 
appear, and $\alpha$ might vary accordingly, as found in \cite{zm}.  
We conclude that the Navier condition is
an appropriate and genuine boundary condition for a liquid-liquid interface.

An outstanding issue is the value of the slip coefficient $\alpha$.
If we use parameter values for Argon (for which the Lennard-Jones potential 
with $A_{ij}=1$ is quantitatively valid) to translate the (1,1)
coefficient into physical units, we have $\alpha \sim 10^{-5}$m/Pa s, a 
value three orders of magnitude larger than observed or
inferred in polymer melts \cite{lam,zm} at low shear rates.  
A likely explanation for the discrepancy is that the interactions used here 
may be too repulsive as compared to those in the experimental systems.  
To pursue this possibility, in the  (2,4) system we ran additional 
simulations with decreasing immiscibility, using a sequence of higher values 
of the inter-liquid interaction strengh $A_{12}=0.2\ldots 0.8$.  The 
apparent slip and the density dip were found to decrease roughly linearly to 
zero from their values at $A_{12}=0$ in Table I.  
More generally, $\alpha$ depends in a non-trivial way on the 
molecular structure and interaction of both fluids present at the interface,
as well as operating conditions such as temperature and density, and perhaps
on driving force as well at higher shear and velocity, and little insight 
into its value is available at the moment.

A dip in the density at a liquid-liquid interface is somewhat unusual, and
in the light of the preceding paragraph one may be concerned about the
realism of the Lennard-Jones potentials used in this paper.  In fact, 
simulations in the literature using fairly realistic interactions either
do or do not exhibit a dip, depending on the liquids involved:  for example,
a dip is present at the water/octane interface \cite{octane} but not in 
the water/carbon tetrachloride case \cite{ccl4}.  Experimental evidence for
a density dip is lacking, but an experimental measurement is difficult.
While it is possible to obtain high resolution normal to an interface using
x-ray scattering for example, the horizontal resolution is much coarser, and
at larger length scales an interface is subject to thermal roughening which
would smooth the density profile.

We thank M.\ M.\ Denn and M.\ Rauscher for discussions.
This work was supported by the NASA Exploration Systems Mission Directorate.

\end{document}